\documentclass[10pt,preprint]{aastex}

\usepackage{graphicx}
\usepackage{times}
\usepackage{amssymb}

\newcommand{\prt}{\partial}
\newcommand{\mrm}{\mathrm}

\begin{document}

\title{Secular instability in quasi-viscous disc accretion}

\author{Jayanta K. Bhattacharjee}
\affil{Department of Theoretical Physics, \\ Indian Association for the
Cultivation of Science, \\Jadavpur, Calcutta (Kolkata) 700032, India}
\email{tpjkb@mahendra.iacs.res.in}

\and

\author{Arnab K. Ray}
\affil{Inter--University Centre for Astronomy and Astrophysics, \\
Post Bag 4, Ganeshkhind, Pune University Campus, \\
Pune 211019, India}
\email{akr@iucaa.ernet.in}

\begin{abstract}
A first-order correction in the $\alpha$-viscosity parameter of Shakura
\& Sunyaev has been introduced in the standard inviscid and thin accretion 
disc.
A linearised time-dependent perturbative study of the stationary solutions
of this ``quasi-viscous" disc leads to the development of a secular
instability on large spatial scales. This qualitative feature is equally
manifest for two
different types of perturbative treatment --- a standing wave on subsonic
scales, as well as a radially propagating wave. Stability of the flow is
restored when viscosity disappears.
\end{abstract}


\section{Introduction}
\label{sec1}

The compressible, inviscid and thin disc flow has by now become an  
established model in accretion studies~\citep{az81,fuk87,c89,nf89, 
skc90,ky94,yk95,par96,msc96,lyyy97,das02,dpm03,ray03a,bdw04,das04,
abd06,dbd06,crd06,gkrd07}. This is a particularly
expedient and simple physical system to study, especially as regards
the rotating flow in the innermost regions of the disc, in the
vicinity of a black hole. Steady global solutions of inviscid 
axisymmetric accretion on to a black hole have been meticulously
studied over the years, and at present there exists an extensive
body of literature devoted to the subject, with especial emphasis
on the transonic nature of solutions, the multitransonic character
of the flow, formation of shocks, and the stability of global solutions 
under time-dependent linearised perturbations. 

Having stressed the usefulness of the inviscid model among researchers
in accretion astrophysics, it must also be recognised that this model 
has its own limitations. 
It is easy to understand that while the presence of angular momentum leads 
to the formation of an accretion disc in the first place, a physical 
mechanism must also be found for the outward transport of angular 
momentum, which should then make possible the inward drift of the accreting 
matter into the potential well of the accretor. Viscosity has been known 
all along to be a just such a physical means to effect infall, although
the exact presciption for viscosity in an accretion disc is still
a matter of much debate~\citep{fkr02}. What is well appreciated, however,
is that the viscous prescription should be compatible with an enhanced
outward transport of angular momentum. The very well-known $\alpha$
parametrisation of~\citet{ss73} is based on this principle. 

And so it transpires that on global scales --- especially on the
very largest scales of the disc --- the inviscid model will encounter
difficulties in the face of the fact that without an effective 
outward transport of angular momentum, the accretion process cannot 
be sustained globally. To address this adverse issue, what is being 
introduced in this paper is the ``quasi-viscous" disc model. This model 
involves prescribing a very small first-order viscous correction in the 
$\alpha$-viscosity parameter of~\citet{ss73}, about the zeroth-order
inviscid solution. In doing this, a viscous generalisation of the 
inviscid flow can be logically extended to capture the important physical
properties of accretion discs on large length scales, without compromising 
on the fundamentally simple and elegant features of the inviscid model. 
This is the single most appealing aspect of the quasi-viscous disc model 
vis-a-vis many other standard models of axisymmetric flows which 
involve viscosity~\citep{ss73,lt80,pri81,mkfo84,bmc86,acls88,ny94,
c96a,c96b,c96c,cal97,pa97,fkr02,ap03,unrs06}. 

While many previous works have also taken up the question of the
stability of viscous thin disc 
accretion~\citep{le74,ss76,ls77,kato78,us05}, 
the specific objective of the present paper is to study the stability 
of stationary 
quasi-viscous inflow solutions under the influence of a time-dependent 
and linearised radial perturbation. It has already been shown in some
earlier works that inviscid solutions are stable under a perturbation 
of this kind~\citep{ray03a,crd06}. 
What has been found through this particular work is that with the merest 
presence of viscosity (i.e. to a first order in $\alpha$, which itself 
is much less than unity) about the stationary inviscid solutions, 
instabilities develop exponentially on large length scales. This has 
disturbing implications, because all physically meaningful inflow solutions 
will have to pass through these length scales, connecting the outer 
boundary of the flow with the surface of the accretor (or the event 
horizon, if the accretor is a black hole). 

The perturbative treatment has been executed in two separate ways --- 
as a standing wave and as a high-frequency travelling wave, and in both
cases the perturbation displays growth behaviour. It need not always be
true that standing and travelling waves will simultaneously exhibit the
same qualitative properties as far as stability is concerned. Many 
instances in fluid dynamics bear this out. In the case of binary 
fluids, standing waves indicate instability as opposed to travelling
waves~\citep{ch93,bb05}, while the whole physical picture is quite 
the opposite for the fluid dynamical problem of the hydraulic 
jump~\citep{bdp93,rb04,sbr05}. Contrary to all this the axisymmetric
stationary quasi-viscous flow is greatly disturbed both by a standing 
wave and by a travelling wave. This provides convincing evidence of 
its unstable character, and it is very much in consonance with similar 
conclusions drawn from some earlier studies. For high-frequency radial
perturbations~\citet{ct93} have found that inertial-acoustic modes are
locally unstable, with a greater degree of growth for the outward 
travelling modes than the inward ones. On the other hand,~\citet{kat88} 
have revealed a growth in the amplitude of a non-propagating perturbation 
at the critical point, which, however, stabilises in the inviscid regime. 

This kind of instability --- one that manifests itself only if 
some dissipative mechanism (viscous dissipation in the case of the 
quasi-viscous rotational flow) is operative --- is called 
{\em secular instability}~\citep{sc87}. It should be very much instructive 
here to furnish a parallel instance of the destabilising influence of 
viscous dissipation in a system undergoing rotation : that of the effect 
of viscous dissipation in a Maclaurin spheroid~\citep{sc87}.
In studying ellipsoidal figures of equilibrium,~\citet{sc87}
has discussed that a secular instability develops in a
Maclaurin spheroid, when the stresses derive from an ordinary viscosity
which is defined in terms of a coefficient of kinematic viscosity (as
the $\alpha$ parametrisation is for an accretion disc), and when the 
effects arising from viscous dissipation are considered as small 
perturbations on the inviscid flow, to be taken into account in the 
first order only. It is exactly in this spirit that the ``quasi-viscous" 
approximation has been prescribed for the thin accretion disc, although,
unlike a Maclaurin spheroid, an astrophysical accretion disc is an open
system.

Curiously enough, the geometry of the fluid flow also seems to be 
having a bearing on the issue of stability. The same kind of
study, as has been done here with viscosity in a rotational flow, had
also been done earlier for a viscous spherically symmetric accreting
system. In that treatment~\citep{ray03b} viscosity was found to have 
a stabilising influence on the system, causing a viscosity-dependent 
decay in the amplitude of a linearised standing-wave perturbation. 
This is quite in keeping with the understanding that the respective
roles of viscosity are at variance with each other in the two 
distinctly separate cases of spherically symmetric flows and thin
disc flows. While viscosity contributes to the resistance against 
infall in the former case, in the latter it aids the infall process. 

Finally, an important aspect of the time-dependent 
perturbative analysis that
may be emphasised is that although the flow has been considered to
be driven by the Newtonian potential, none of the physical conclusions
of this work will be qualified in any serious way upon using any of the 
pseudo-Newtonian potentials~\citep{pw80,nw91,abn96}, which are regularly 
invoked in accretion-related literature to describe rotational flows 
on to a Schwarzschild black hole, even while preserving the Newtonian 
construct of space and time. This shall be especially true 
of the flow on large scales, where all pseudo-Newtonian potentials 
converge to the Newtonian limit, and, therefore, the conclusions of 
this perturbative treatment will also have a similar bearing on 
pseudo-Schwarzschild flows. 

\section{The quasi-viscous axisymmetric flow}
\label{sec2}

For the thin disc, under the condition of hydrostatic equilibrium along 
the vertical direction~\citep{mkfo84,fkr02}, two of the relevant flow 
variables are the drift velocity, $v$, and the surface density, $\Sigma$. 
In the thin-disc approximation the latter has been defined by vertically 
integrating the volume density, $\rho$, over the disc 
thickness, $H(r)$. This gives $\Sigma \cong \rho H$, and in terms 
of $\Sigma$, the continuity equation is set down as 
\begin{equation}
\label{consig}
\frac{\prt \Sigma}{\prt t}+ \frac{1}{r} \frac{\prt}{\prt r}
\left(\Sigma vr \right) =0 .
\end{equation}
For a flow driven by the Newtonian potential, $V(r) = -GMr^{-1}$,
assumption of the 
hydrostatic equilibrium in the vertical direction will give the 
condition $H = r (c_{\mrm s}/v_{\mrm K})$, in which the local speed
of sound, $c_{\mrm s}$, and the local Keplerian velocity, $v_{\mrm K}$, 
are, respectively, defined as 
$c_{\mrm s}^2 = \gamma K \rho^{\gamma -1}$ and 
$v_{\mrm K}^2 = GMr^{-1}$, with the constants $\gamma$ and $K$
deriving from the application of a polytropic equation of state, 
$P= K \rho^{\gamma}$, in terms of which, the speed of sound may 
given as $c_{\mrm s}^2 = \prt P/\prt \rho$. Written explicitly, the 
disc height is, therefore, expressed as 
\begin{equation}
\label{aitch}
H = \left(\frac{\gamma K}{GM} \right)^{1/2} \rho^{(\gamma -1)/2}
r^{3/2} ,
\end{equation}
and with the use of this result, the continuity equation could then
be recast as 
\begin{equation}
\label{conrho}
\frac{\prt}{\prt t} \left[\rho^{(\gamma +1)/2}\right] 
+\frac{1}{r^{5/2}} 
\frac{\prt}{\prt r} \left[\rho^{(\gamma +1)/2}vr^{5/2}\right]=0 .
\end{equation}

The condition for the balance of specific angular momentum in the flow
is given by~\citep{fkr02}, 
\begin{equation}
\label{angsig}
\frac{\prt}{\prt t}\left(\Sigma r^2 \Omega \right) + \frac{1}{r}
\frac{\prt}{\prt r} \left[ \left(\Sigma v r \right) r^2 \Omega \right]
= \frac{1}{2 \pi r} \left(\frac{\prt \mathcal{G}}{\prt r}\right) ,
\end{equation}
where $\Omega$ is the local angular velocity of the flow, while the 
torque is given as 
\begin{equation}
\label{torque}
{\mathcal G} = 2 \pi r \nu \Sigma r^2 
\left(\frac{\prt \Omega}{\prt r}\right) ,
\end{equation}
with $\nu$ being the kinematic viscosity associated with the flow. With 
the use of the continuity equation, as equation~(\ref{consig}) gives it,
and going by the~\citet{ss73} prescription for the kinematic viscosity,
$\nu = \alpha c_{\mrm s} H$, it would be easy to reduce 
equation~(\ref{angsig}) to the form~\citep{fkr02,ny94} 
\begin{equation}
\label{angrho}
\frac{1}{v}\frac{\prt}{\prt t}\left(r^2 \Omega \right)
+ \frac{\prt}{\prt r}\left(r^2 \Omega \right) = \frac{1}{\rho vrH}
\frac{\prt}{\prt r}\left[\frac{\alpha \rho H c_{\mrm s}^2 r^3}
{\Omega_{\mrm K}}\left(\frac{\prt \Omega}{\prt r}\right)\right] ,
\end{equation}
with $\Omega_{\mrm K}$ being defined from $v_{\mrm K}= r \Omega_{\mrm K}$. 

Going back to equation~(\ref{conrho}), a new variable is defined as 
$f= \rho^{(\gamma +1)/2}vr^{5/2}$, whose steady value, as it is very easy 
to see from equation~(\ref{conrho}), can be closely identified with the 
constant matter flux rate. In terms of this new variable,
equation~(\ref{conrho}) can be modified as 
\begin{equation}
\label{coneff}
\frac{\prt}{\prt t} \left[\rho^{(\gamma +1)/2}\right]
+\frac{1}{r^{5/2}}
\frac{\prt f}{\prt r} =0 ,
\end{equation}
while equation~(\ref{angrho}) can be rendered as 
\begin{equation}
\label{angeff}
\frac{1}{v}\frac{\prt}{\prt t}\left(r^2 \Omega \right)
+ \frac{\prt}{\prt r}\left(r^2 \Omega \right) = \alpha 
\left(\frac{\gamma K}{GM} \right) \frac{1}{f}
\frac{\prt}{\prt r}\left[f \left(\frac{f^2 \Omega_{\mrm K}}
{\rho^2 v^3} \right) \frac{\prt \Omega}{\prt r}\right] .
\end{equation}

\begin{figure}
\begin{center}
\includegraphics[angle=90,scale=0.6]{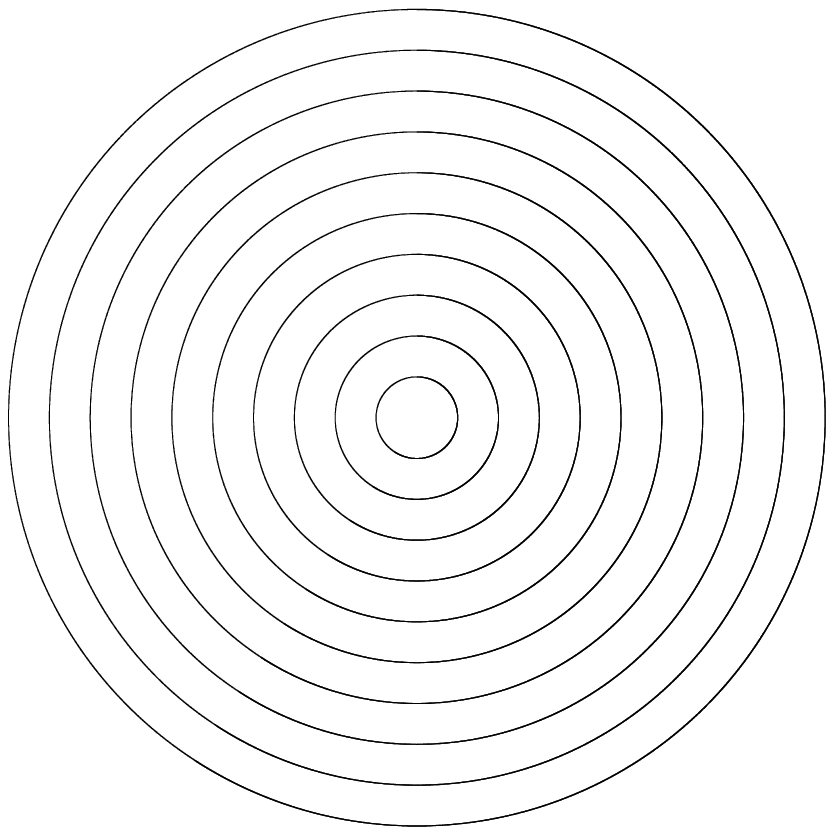}
\caption{A schematic representation of the axisymmetric inviscid disc.
Viewed from the top (along the vertical axis of the disc) the flow will
be seen to form closed circular paths.}
\label{f1}
\end{center}
\end{figure}

\begin{figure}
\begin{center}
\includegraphics[angle=90,scale=0.6]{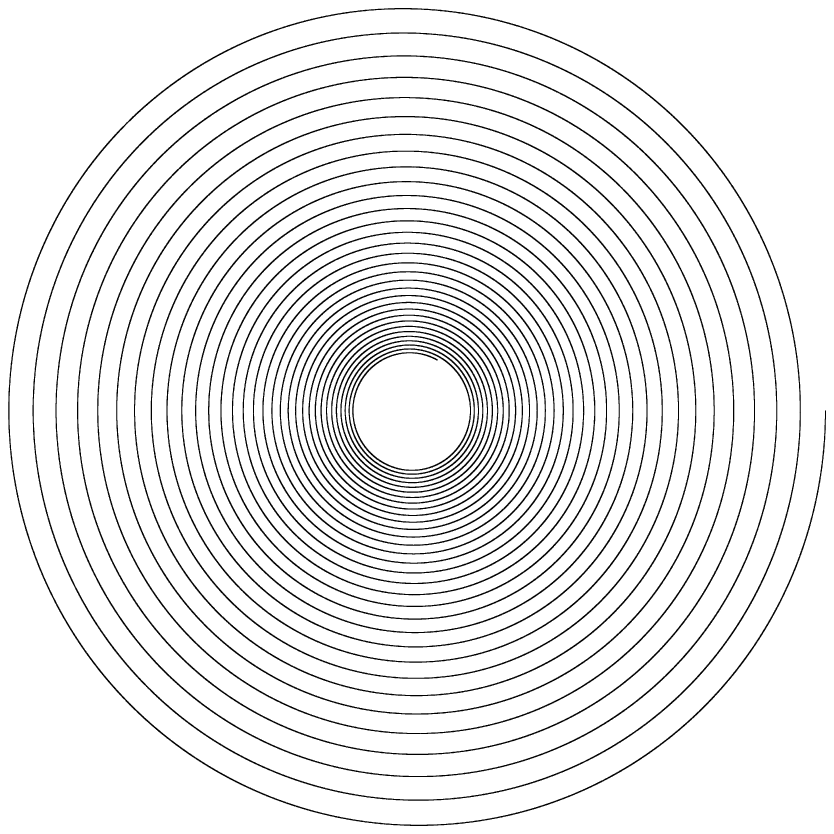}
\caption{A schematic representation to trace the spiralling
behaviour of the physical flow trajectories upon the introduction
of a small viscous correction to the background inviscid flow.}
\label{f2}
\end{center}
\end{figure}

The inviscid disc model is given by the requirement that 
$r^2 \Omega = \lambda$, in which $\lambda$ is the constant specific
angular momentum. The quasi-viscous disc that is being proposed 
here will introduce a first-order correction in terms involving 
$\alpha$, the~\citet{ss73} viscosity parameter, about the 
constant angular momentum solution. Mathematically this will be  
represented by the prescription of an effective specific angular
momentum,
\begin{equation}
\label{effecang}
\lambda_{\mrm{eff}}(r) = r^2 \Omega = \lambda + \alpha r^2 
\tilde{\Omega} ,
\end{equation}
with the form of $\tilde{\Omega}$ having to be determined from 
equation~(\ref{angeff}), under the stipulation that the dimensionless 
$\alpha$-viscosity parameter is much smaller than unity. This smallness 
of the quasi-viscous correction induces only very small changes on the 
constant angular momentum background, and, therefore, neglecting all 
orders of $\alpha$ higher than the first, and ignoring any explicit 
time-variation of the viscous correction term, the latter being a 
standard method adopted also for Keplerian 
flows~\citep{le74,ss76,pri81,fkr02}, 
the dependence of $\tilde{\Omega}$ on $v$ and $\rho$ is obtained as 
\begin{equation}
\label{tilomeg}
\tilde{\Omega} = - \frac{2 \lambda}{r^2} \left(\frac{\gamma K}
{GM}\right)\left[\frac{f^2 \Omega_{\mrm K}}{\rho^2 v^3 r^3} +
\int \frac{f^2 \Omega_{\mrm K}}{\rho^2 v^3 r^3} \left(\frac{1}{f}
\frac{\prt f}{\prt r}\right)\, {\mrm d}r \right] . 
\end{equation}
The effect of the absence and the presence of a small viscous 
correction to the inviscid background flow has been schematically
shown in Figs.~\ref{f1} \&~\ref{f2}, respectively. 

The stationary solution of equation~(\ref{coneff}) can be easily
obtained as a first integral, and with the help of equation~(\ref{aitch}),
it can be set down as 
\begin{equation}
\label{statiocon}
2 \pi \left(\frac{\gamma K}{GM}\right)^{1/2} \rho^{(\gamma +1)/2} 
v r^{5/2} = -\dot{m} , 
\end{equation}
with $\dot{m}$ being the conserved matter inflow rate. The negative sign
arises because for inflows, $v$ goes with a negative sign. Further, under
stationary conditions, equation~(\ref{effecang}) can be written in a
modified form as 
\begin{equation}
\label{statioang}
\lambda_{\mrm{eff}}(r) = \lambda - 2 \alpha \lambda 
\left(\frac{c_{\mrm s}^2}{vv_{\mrm K}} \right) .  
\end{equation}
From equation~(\ref{statiocon}), with $\rho$ approaching a constant
asymptotic value on large length scales, the drift velocity, $v$, can
be seen to go asymptotically as $r^{-5/2}$. Bearing in mind that for
inflows, $v<0$, the asymptotic dependence of the effective angular
momentum can be shown to be  
\begin{equation}
\label{asympang}
\lambda_{\mrm{eff}}(r) \sim \lambda + 2 \alpha \lambda 
\left(\frac{r}{r_{\mrm s}}\right)^3 , 
\end{equation}
with $r_{\mrm s}$ being a scale of length, which, to an
order-of-magnitude, is given by 
$r_{\mrm s}^3\sim GM{\dot{m}}[c_{\mrm s}^3(\infty)\rho(\infty)]^{-1}$. 
This asymptotic behaviour is entirely to be expected, because the 
physical role of viscosity is to transport angular momentum to large
length scales of the accretion disc. 

Lastly, the equation for radial momentum balance in the flow will 
also have to be modified under the condition of quasi-viscous 
dissipation. This has to be done according to the scheme outlined
in equation~(\ref{effecang})
by which, the centrifugal term, $\lambda^2_{\mrm{eff}}(r)/r^3$, of 
the radial momentum balance equation, will have to be corrected 
upto a first order in $\alpha$. This will finally lead to the result
\begin{equation}
\label{radmom}
\frac{\prt v}{\prt t} + v \frac{\prt v}{\prt r}
+ \frac{1}{\rho}\frac{\prt P}{\prt r} + V^{\prime}(r)
- \frac{\lambda^2}{r^3} - 2 \alpha \frac{\lambda}{r^3}
\left(r^2 \tilde{\Omega} \right) =0 , 
\end{equation}
with $\tilde{\Omega}$ being given by equation~(\ref{tilomeg}), and $P$ 
being expressed as a function of $\rho$ with the help of a polytropic 
equation of state, as has been mentioned earlier. The steady solution
of equation~(\ref{radmom}) is given as 
\begin{equation}
\label{statiorad}
v \frac{{\mrm d} v}{{\mrm d} r} + \frac{1}{\rho}
\frac{{\mrm d}P}{{\mrm d} r} + V^{\prime}(r) - \frac{\lambda^2}{r^3}
+ 4 \alpha \frac{\lambda^2}{r^3} 
\left(\frac{c_{\mrm s}^2}{vv_{\mrm K}} \right) = 0 ,
\end{equation}
whose first integral cannot be obtained analytically because of the
$\alpha$-dependent term. In the inviscid limit, though, the integral
is easily obtained. This case will be governed by conserved conditions, 
and its solutions have been well-known in accretion 
literature~\citep{c89,das02,dpm03}. They will either be open solutions
passing through saddle points or closed paths about centre-type points. 
The slightest presence of viscous dissipation, however, will radically
alter the nature of solutions seen in the inviscid limit, and it may be 
easily understood that solutions forming closed paths about centre-type
points, will, under conditions of small-viscous correction, be of the 
spiralling kind~\citep{lt80,mkfo84,ap03}. This state of affairs is 
appreciated very easily by the analogy of the simple harmonic oscillator.
In the undamped state the phase trajectories of the oscillator will,
very much like the solutions of the inviscid flow, be either closed 
paths about centre-type points or open paths through saddle points. 
With the presence of even very weak damping the closed paths change 
into spiralling solutions. 

After having understood the qualitative nature of the stationary 
flows in the quasi-viscous accretion disc, as given by 
equations~(\ref{statiocon}),~(\ref{statioang}) 
and~(\ref{statiorad}), it will now become possible to carry out a 
real-time linear stability analysis about the stationary solutions, 
with the help of equations~(\ref{conrho}) and~(\ref{radmom}).

\section{An equation for a time-dependent perturbation on stationary
solutions}
\label{sec3}

About the stationary solutions of the flow variables, $v$ and $\rho$,
a time-dependent perturbation is introduced according to the scheme,
$v(r,t) = v_0(r) + v^{\prime}(r,t)$, $\rho (r,t) = \rho_0(r) + 
\rho^{\prime}(r,t)$ and $f(r,t) = f_0(r) + f^{\prime}(r,t)$, in all
of which, the subscript ``$0$" implies stationary values, with $f_0$
especially, as can be seen from equation~(\ref{conrho}), being a constant. 
This constant, as it is immediately evident from a look at 
equation~(\ref{statiocon}), is very much connected to the matter flow
rate, and, therefore, the perturbation $f^{\prime}$ is to be seen as 
a disturbance on the steady, constant background accretion rate. 
For spherically symmetric flows, this Eulerian perturbation scheme has 
been applied by~\citet{pso80} and~\citet{td92}, while for inviscid 
axisymmetric flows, the same method has been used equally effectively 
by~\citet{ray03a,ray03b} and~\citet{crd06}.   

The definition of $f$ will lead to a linearised dependence among 
$f^{\prime}$, $v^{\prime}$ and $\rho^{\prime}$ as 
\begin{equation}
\label{effprime}
\frac{f^{\prime}}{f_0} = \left( \frac{\gamma +1}{2} \right)
\frac{\rho^{\prime}}
{\rho_0} + \frac{v^{\prime}}{v_0} ,
\end{equation}
while from equation~(\ref{conrho}), an exclusive dependence 
of $\rho^{\prime}$
on $f^{\prime}$ will be obtained as 
\begin{equation}
\label{flucden}
\frac{\prt \rho^{\prime}}{\prt t} + \beta^2
\frac{v_0 \rho_0}{f_0} \left(\frac{\prt f^{\prime}}{\prt r}\right)=0
\end{equation}
with $\beta^2 = 2(\gamma +1)^{-1}$. 
Combining equations~(\ref{effprime}) and~(\ref{flucden}) will 
render the velocity fluctuations as
\begin{equation}
\label{flucvel}
\frac{\prt v^{\prime}}{\prt t}= \frac{v_0}{f_0}
\left(\frac{\prt f^{\prime}}{\prt t}+{v_0}
\frac{\prt f^{\prime}}{\prt r}\right)
\end{equation}
which, upon a further partial differentiation with respect to time,
will give
\begin{equation}
\label{flucvelder2}
\frac{{\prt}^2 v^{\prime}}{\prt t^2}= \frac{v_0}{f_0}
\left[\frac{\prt^2 f^{\prime}}{\prt t^2} + v_0 \frac{\prt}{\prt r}
\left(\frac{\prt f^{\prime}}{\prt t}\right) \right] . 
\end{equation}

From equation~(\ref{radmom}) the linearised fluctuating part could be
extracted as
\begin{equation}
\label{flucradmom}
\frac{\prt v^{\prime}}{\prt t}+ \frac{\prt}{\prt r}
\left( v_0 v^{\prime} + c_{\mrm{s0}}^2
\frac{\rho^{\prime}}{\rho_0}\right)+4\alpha \lambda^2 \frac{\sigma}{r^3}
\left[2 \frac{f^{\prime}}{f_0} - 2 \frac{\rho^{\prime}}{\rho_0} 
-3 \frac{v^{\prime}}{v_0} + \frac{1}{\sigma} \int \sigma 
\frac{\prt}{\prt r} \left(\frac{f^{\prime}}{f_0}\right)\, {\mrm d}r
\right] =0 , 
\end{equation}
in which $\sigma = c_{\mrm{s0}}^2/(v_0 v_{\mrm K})$ and $c_{\mrm{s0}}$ 
is the local speed of sound in the steady state. Differentiating 
equation~(\ref{flucradmom}) partially with respect to $t$,
and making use of equations~(\ref{flucden}), (\ref{flucvel}) 
and~(\ref{flucvelder2}) to substitute for all the first and second-order
derivatives of $v^{\prime}$ and $\rho^{\prime}$, will deliver the result
\begin{displaymath}
\label{tpert1}
\frac{{\prt}^2 f^{\prime}}{\prt t^2} +2 \frac{\prt}{\prt r}
\left(v_0 \frac{\prt f^{\prime}}{\prt t} \right) + \frac{1}{v_0}
\frac{\prt}{\prt r}\left[ v_0 \left(v_0^2-
\beta^2 c_{\mrm{s0}}^2 \right) \frac{\prt f^{\prime}}{\prt r}\right] 
- 4 \alpha \lambda^2 \frac{\sigma}{v_0 r^3}\bigg[\frac{\prt f^{\prime}}
{\prt t} + \left(\frac{3\gamma -1}{\gamma + 1}\right) v_0 
\frac{\prt f^{\prime}}{\prt r} 
\end{displaymath}
\begin{equation}
\label{tpert}
\qquad \qquad \qquad \qquad \qquad \qquad \qquad \qquad \qquad \qquad
\qquad \qquad \qquad \qquad \qquad \qquad
- \frac{1}{\sigma} \int \sigma 
\frac{\prt}{\prt r} \left(\frac{\prt f^{\prime}}{\prt t} \right)\, 
{\mrm d}r \bigg] = 0 
\end{equation}
entirely in terms of $f^{\prime}$. This is the equation of motion for
a perturbation imposed on the constant mass flux rate, $f_0$, and it 
shall be important to note here that the choice of a driving potential,
Newtonian or pseudo-Newtonian, has no explicit bearing on the form 
of the equation. 

\section{Linear stability analysis of stationary solutions}
\label{sec4}

With a linearised equation of motion for the perturbation having been
derived, a solution of the form 
$f^{\prime}(r,t) = g_{\omega}(r) \exp(-{\mrm i}\omega t)$ is applied 
to it. From equation~(\ref{tpert}), this will give
\begin{displaymath}
\label{disp11}
\omega^2 g_\omega + 2 {\mrm i} \omega \frac{\mrm d}{{\mrm d}r}
\left(v_0 g_\omega \right)
- \frac{1}{v_0} \frac{\mrm d}{{\mrm d}r}
\left[v_0 \left(v_0^2 - \beta^2 c_{\mrm{s0}}^2 \right)
\frac{{\mrm d}g_\omega}{{\mrm d}r} \right] + 4 \alpha \lambda^2 
\frac{\sigma}{v_0 r^3} \bigg[ - {\mrm i}\omega g_\omega + 
\left(\frac{3 \gamma -1}{\gamma + 1} \right) v_0 
\frac{{\mrm d}g_\omega}{{\mrm d}r} 
\end{displaymath}
\begin{equation}
\label{disp1}
\qquad \qquad \qquad \qquad \qquad \qquad \qquad \qquad \qquad \qquad
\qquad \qquad \qquad \qquad \qquad \qquad
+ \frac{{\mrm i} \omega}{\sigma}
\int \sigma \left(\frac{{\mrm d}g_\omega}{{\mrm d}r}\right) \, 
{\mrm d}r \bigg] = 0 .
\end{equation}
The perturbation may now be treated as a standing wave spatially
confined between two suitably chosen boundaries, as well as a radially
propagating high-frequency wave. These two distinct cases will be 
taken up separately
in what follows, to see how stationary solutions are affected by the 
perturbation. 

\subsection{Standing waves}
\label{subsec41}

It can be easily appreciated that with viscous dissipation present 
in the accreting system, multiply-valued solutions in the phase 
portrait are a distinct possibility~\citep{lt80,ak89,ap03}. This, 
however, is physically not feasible for a fluid flow, and, therefore, 
for all solutions which are multiply-valued about a critical point,  
it should be necessary to have the inner branch 
of a solution fitted to its outer branch via a shock, which is a standard
practice in general fluid dynamical studies~\citep{bdp93}. The outer 
branch of this discontinuous solution will connect the shock with the 
outer boundary of the disc itself, and in the phase portrait of the 
flow, there will exist an entire family of these outer solutions which 
arguably shall be subsonic. It would be worthwhile at
this point to remember that for this kind of a thin disc system, the 
local speed of acoustic propagation would go as $\beta c_{\mrm{s0}}$,
and subsonic solutions would have their bulk flow velocity, $v_0$, at 
less than this value. Over the entire subsonic range, at two chosen points,
the perturbation may be spatially constrained by requiring it to be 
a standing wave, which dies out at the two chosen boundaries. The outer
point could suitably be chosen to be at the outer boundary of the flow
itself, where, by virtue of the boundary condition on the steady flow,
the perturbation would naturally decay out. The inner boundary of the
standing wave is to be chosen infinitesimally close to the shock front,
through which the flow will be discontinuous, and the perturbation
will be made to die out in its neighbourhood. Between these two points, for 
solutions which are entirely subsonic, it should be necessary to multiply 
equation~(\ref{disp1}) by $v_0 g_\omega$, and then carry out an integration
by parts. The requirement that all integrated ``surface terms" vanish
at the two boundaries of the standing wave, will give a quadratic
dispersion relation in $\omega$, which will read as 
\begin{equation}
\label{disp2}
{\mathcal A} \omega^2 + {\mathcal B} \omega + {\mathcal C} = 0 ,
\end{equation}
in which the coefficients of each successive term will be given by 
\begin{displaymath}
\label{coeff12}
{\mathcal A}= \int \left(v_0 g^2_\omega \right) \, {\mrm d}r , 
\qquad \qquad 
{\mathcal B}=-4{\mrm i}\alpha\lambda^2\int\left[\frac{g_\omega}{r^3} 
\int g_\omega \left(\frac{{\mrm d}\sigma}{{\mrm d}r}\right)
\,{\mrm d}r\right]\,{\mrm d}r  
\end{displaymath}
and
\begin{displaymath}
\label{coeff3}
{\mathcal C} = \int v_0 \left(v_0^2 - \beta^2 c_{\mrm{s0}}^2 \right)
\left(\frac{{\mrm d}g_\omega}{{\mrm d}r}\right)^2\,{\mrm d}r + 2 \alpha 
\lambda^2 \left(3 \gamma -1 \right) \int 
\frac{\beta^2 \sigma g_\omega v_0}{r^3}
\left(\frac{{\mrm d}g_\omega}{{\mrm d}r}\right)\,{\mrm d}r . 
\end{displaymath}

It is possible now to find a solution for $\omega$, but of greater 
immediate significance is the fact that this solution will yield an
$\alpha$-dependent real part of the temporal component of the perturbation,
which will go as 
\begin{equation}
\label{real}
\Re \left(- {\mrm i} \omega \right) = 2 \alpha
\left[\int \left(v_0 g^2_\omega \right) \xi (r) \, {\mrm d}r \right]
\left[ \int \left(v_0 g^2_\omega \right) \, {\mrm d}r \right]^{-1} 
\sim \alpha \xi (r) ,
\end{equation} 
with $\xi (r)$ itself being expressed as
\begin{equation}
\label{xi}
\xi (r) = \frac{\lambda^2}{v_0 g_\omega r^3} \int g_\omega 
\left(\frac{{\mrm d}\sigma}{{\mrm d}r} \right) \, {\mrm d}r \, 
\sim \frac{\lambda^2 c_{\mrm{s0}}^2}{v_0^2 v_{\mrm K} r^3} 
= \frac{\lambda^2}{{\mathcal M}^2 v_{\mrm K} r^3} , 
\end{equation}
in which the Mach number, ${\mathcal M} = v_0/c_{\mrm{s0}}$. It is 
globally true that $v_{\mrm K} \sim r^{-1/2}$. In this situation, 
under the assumption that the Mach number has a power-law dependence
on the radial distance, given as ${\mathcal M}^2 \sim r^{-\varepsilon}$, 
it is possible to write down, 
\begin{equation}
\label{varxi}
\frac{{\mrm d}(\ln \xi)}{{\mrm d}(\ln r)} =  
\varepsilon  -\frac{5}{2}. 
\end{equation}

On large length scales, with $c_{\mrm{s0}}$ approaching a constant
ambient value, and with $v_0 \sim r^{-5/2}$ from the continuity 
condition, it is easily deduced that $\xi (r) \sim r^{5/2}$. This 
indicates that on large length scales, the amplitude of the spatially
constrained standing-wave perturbation grows in time, i.e. the subsonic
solutions on which this perturbation has been imposed, display unstable
behaviour. It is easy to see that when $\alpha$ vanishes, all stationary
solutions restricted by the condition $v_0 < \beta c_{\mrm{s0}}$, are 
stable under a standing-wave perturbation, which displays a purely 
oscillatory behaviour with no growth in amplitude. For the inviscid 
disc (with $\alpha = 0$) this has been shown by~\citet{ray03a}. 
Regarding this point, it will be instructive to 
mention again that~\citet{kat88} have pointed to the existence of a 
non-propagating growing perturbation localised at the critical point, 
but interestingly enough they have also found that this instability 
disappears in inviscid transonic flows.

\subsection{Radially propagating waves}
\label{subsec42}

The perturbation is now made to behave in the manner of a radially
travelling wave, whose wavelength is suitably constrained to be small, 
i.e. it is to be smaller than any characteristic length scale in the 
system. A perturbative treatment of this nature has been carried out 
before on spherically symmetric flows~\citep{pso80} and on axisymmetric 
flows~\citep{ray03a,crd06}. In both these cases the radius
of the accretor was chosen as the characteristic length scale in
question, and the wavelength of the perturbation was required to be
much smaller than this length scale. In this study of an axisymmetric 
flow driven by a Newtonian potential, the radius of the accreting star,
$r_\star$, could be a choice for such a length scale. As a result, the 
frequency, $\omega$, of the waves should be large. 

An algebraic rearrangement of terms in equation~(\ref{disp1}) will 
lead to an integro-differential equation of the form 
\begin{equation}
\label{gee}
{\mathcal P}\frac{{\mrm d^2} g_\omega}{{\mrm d}r^2} 
+ {\mathcal Q}\frac{{\mrm d} g_\omega}{{\mrm d}r}
- {\mathcal R}g_\omega 
+ {\mathcal T} \int g_\omega
\left(\frac{{\mrm d} \sigma}{{\mrm d}r}\right)\,{\mrm d}r =0 , 
\end{equation}
with its coefficients being given by 
\begin{displaymath}
\label{peequ}
{\mathcal P} = v_0^2 - \beta^2 c_{\mrm{s0}}^2 , \qquad \qquad 
{\mathcal Q} = 3 v_0 \frac{{\mrm d}v_0}{{\mrm d}r}-\frac{1}{v_0}
{\frac{\mrm d}{{\mrm d}r}}\left( v_0 \beta^2 c_{\mrm{s0}}^2\right)
- 2{\mrm{i}}\omega v_0- 2\alpha \lambda^2\left(3\gamma-1 \right) 
\frac{\beta^2 \sigma}{r^3} ,  
\end{displaymath}
\begin{displaymath}
\label{aartee} 
{\mathcal R} = 2{\mrm i} \omega \frac{{\mrm d}v_0}{{\mrm d}r}
+ \omega^2  \qquad {\mathrm{and}} \qquad
{\mathcal T} = \frac{4 {\mrm{i}} \omega \alpha \lambda^2}{v_0 r^3} . 
\end{displaymath}

At this stage, bearing in mind the constraint that $\omega$ is large,
the spatial part of the perturbation, $g_\omega (r)$, is prescribed as
$g_\omega (r) = \exp(s)$, with the function $s$ itself being represented 
as a power series of the form
\begin{equation}
\label{pow}
s(r)=\sum_{n=-1}^{\infty}\frac{k_n(r)}{\omega^n} . 
\end{equation}
The integral term in equation~(\ref{gee}), can,
through some suitable algebraic substitutions, be recast as 
\begin{displaymath}
\label{integ}
\int g_\omega
\left(\frac{{\mrm d} \sigma}{{\mrm d}r}\right)\, {\mrm d}r
= \int \exp(s)
\left(\frac{{\mrm d} \sigma}{{\mrm d}s}\right)\, {\mrm d}s
= g_\omega (r) {\mathcal S} ,
\end{displaymath}
with $\mathcal{S}$ itself being given by another power series as 
\begin{equation}
\label{powess}
{\mathcal S}= \sum_{m= 1}^{\infty} \left( -1 \right)^{m+ 1} 
\frac{{\mrm d}^m \sigma}{{\mrm d}s^m} . 
\end{equation}
Following this, all the terms in equation~(\ref{gee}) can be expanded with 
the help of the power series for $g_\omega (r)$. Under the assumption 
(whose self-consistency will be justified soon) that to a leading order 
\begin{displaymath}
\label{ess}
{\mathcal S} \sim \frac{{\mrm d}\sigma}{{\mrm d}s}
\simeq \frac{{\mrm d}\sigma}{{\mrm d}r}
\left(\omega \frac{{\mrm d} k_{-1}}{{\mrm d}r}\right)^{-1} , 
\end{displaymath}
the three successive highest-order terms (in a decreasing order) involving 
$\omega$ will be obtained as $\omega^2$, $\omega$ and $\omega^0$. The 
coefficients of each of these terms are to be collected first and then 
individually summed up. This is to be followed by setting each of these
sums separately to zero, which will yield for $\omega^2$, $\omega$
and $\omega^0$, respectively, the conditions
\begin{equation}
\label{omegsq}
\left(v_0^2 - \beta^2 c_{\mrm{s0}}^2\right)
\left( \frac{{\mrm d}k_{-1}}{{\mrm d}r} \right)^2
-2 {\mrm i} v_0 \frac{{\mrm d}k_{-1}}{{\mrm d}r} -1 = 0
\end{equation}
\begin{displaymath}
\label{omeg11}
\left(v_0^2 - \beta^2 c_{\mrm{s0}}^2\right)
\left( \frac{{\mrm d}^2 k_{-1}}{{\mrm d}r^2}
+ 2 \frac{{\mrm d}k_{-1}}{{\mrm d}r}
\frac{{\mrm d}k_0}{{\mrm d}r} \right)
+ \left[ 3 v_0 \frac{{\mrm d}v_0}{{\mrm d}r}
- \frac{1}{v_0}\frac{\mrm d}{{\mrm d}r}
\left( v_0 \beta^2 c_{\mrm{s0}}^2 \right) -2 \alpha \lambda^2 
\left(3 \gamma -1 \right) \frac{\beta^2 \sigma}{r^3} \right]
\frac{{\mrm d}k_{-1}}{{\mrm d}r}
\end{displaymath}
\begin{equation}
\label{omeg1}
\qquad \qquad \qquad \qquad \qquad \qquad \qquad \qquad \qquad \qquad 
\qquad \qquad \qquad \qquad \qquad 
- 2{\mrm i} v_0 \frac{{\mrm d}k_0}{{\mrm d}r}
- 2{\mrm i} \frac{{\mrm d}v_0}{{\mrm d}r} = 0
\end{equation}
and
\begin{displaymath}
\label{omeg0}
\left( v_0^2 - \beta^2 c_{\mrm{s0}}^2 \right)
\left[ \frac{{\mrm d}^2 k_0}{{\mrm d}r^2}
+ 2 \frac{{\mrm d}k_{-1}}{{\mrm d}r}
\frac{{\mrm d}k_1}{{\mrm d}r} +
\left( \frac{{\mrm d}k_0}{{\mrm d}r} \right)^2
\right] + 
\left [3 v_0 \frac{{\mrm d}v_0}{{\mrm d}r}
- \frac{1}{v_0} \frac{\mrm d}{{\mrm d}r}
\left(v_0 \beta^2 c_{\mrm{s0}}^2 \right) 
- 2 \alpha \lambda^2 \left(3 \gamma -1\right)\frac{\beta^2 \sigma}
{r^3} \right]\frac{{\mrm d}k_0}{{\mrm d}r}
\end{displaymath}
\begin{equation}
\label{contdomeg0}
\qquad \qquad \qquad \qquad \qquad \qquad
\qquad \qquad \qquad \qquad \qquad \qquad
- 2{\mrm{i}}{v_0}\frac{{\mrm d}k_1}{{\mrm d}r}
+ 4 {\mrm i} \alpha \lambda^2 \frac{1}{v_0 r^3} 
\frac{{\mrm d}\sigma}{{\mrm d}r}
\left(\frac{{\mrm d} k_{-1}}{{\mrm d}r}\right)^{-1} = 0 . 
\end{equation}
Out of these, the first two, i.e. equations~(\ref{omegsq}) 
and~(\ref{omeg1}), will deliver the solutions
\begin{equation}
\label{kayminus1}
k_{-1} = \int \frac{{\mrm{i}}}{v_0 \pm \beta c_{\mrm{s0}}} \,
{\mrm d}r
\end{equation}
and
\begin{equation}
\label{kaynot}
k_0 = - \frac{1}{2} \ln \left( v_0 \beta c_{\mrm{s0}} \right)
\pm \alpha \lambda^2 \left(3 \gamma -1 \right) \int
\frac{\beta c_{\mrm{s0}}\left(v_0 \pm \beta c_{\mrm{s0}}\right)}
{v_0 v_{\mrm{K}} r^3\left(v_0^2 - \beta^2 c_{\mrm{s0}}^2\right)} \,
{\mrm{d}}r , 
\end{equation}
respectively. 

The two foregoing expressions give the leading terms in the power 
series of $g_\omega (r)$. While dwelling on this matter, it will
also be necessary to show that all successive terms of $s(r)$ will
self-consistently follow the condition
$\omega^{-n}\vert k_n(r)\vert \gg \omega^{-(n+1)}\vert k_{n+1}(r)\vert$,
i.e. the power series given by $g_\omega (r)$ will converge very quickly 
with increasing $n$. In the inviscid limit, this requirement can be shown 
to be very much true, considering the behaviour of the first three 
terms in $k_n(r)$ from equations~(\ref{kayminus1}), (\ref{kaynot})
and~(\ref{contdomeg0}). These terms can be shown to go
asymptotically as
$k_{-1} \sim r$, $k_0 \sim \ln r$ and $k_1 \sim r^{-1}$, given the
condition that $v_0 \sim r^{-5/2}$ on large length scales, while
$c_{\mrm{s0}}$ approaches its constant ambient value. With the 
inclusion of viscosity as a physical effect, it can be seen from
equations~(\ref{kayminus1}) and~(\ref{kaynot}), respectively, that while
$k_{-1}$ remains unaffected, $k_0$ acquires an $\alpha$-dependent
term that goes asymptotically as $r$. This in itself is an
indication of the extent to which viscosity might alter the inviscid
conditions. However, since $\alpha$ has been chosen to be very much
less than unity, and since the wavelength of the travelling waves is
also very small, implying that
$\omega \gg (v_0 \pm \beta c_{\mrm{s0}})/ r_\star$,
the self-consistency requirement still holds. Therefore, as far
as gaining a qualitative understanding of the effect of viscosity 
is concerned, it should be quite sufficient to truncate the power
series expansion of $s(r)$ after considering the two leading terms 
only, and with the help of these two, an expression for the 
perturbation may be set down as
\begin{equation}
\label{fpertur}
f^{\prime}(r,t) \simeq \frac{A_\pm}{\sqrt{\beta v_0 c_{\mrm{s0}}}}
\exp \left[ \pm \alpha \lambda^2 \left(3 \gamma -1 \right) \int 
\frac{\beta c_{\mrm{s0}}\left(v_0 \pm \beta c_{\mrm{s0}}\right)}
{v_0 v_{\mrm{K}} r^3\left(v_0^2 - \beta^2 c_{\mrm{s0}}^2\right)} 
\, {\mrm d}r \right] 
\exp \left ( \int \frac{{\mrm i} \omega}{v_0 \pm \beta
c_{\mrm{s0}}} \, {\mrm d}r \right)
e^{-{\mrm i} \omega t} , 
\end{equation}
which should be seen as a linear superposition of two waves with 
arbitrary constants $A_+$ and $A_-$. Both of these two waves move 
with a velocity $\beta c_{\mrm{s0}}$ relative to the fluid,
one against the bulk flow and the other along with it, while the bulk
flow itself has a velocity $v_0$. It should be immediately evident 
that all questions pertaining to the growth or decay in the amplitude of
the perturbation will be crucially decided by the real terms delivered
from $k_0$. The viscosity-dependent term is especially crucial
in this regard. For the choice of the lower sign in the real part of
$f^\prime$ in equation~(\ref{fpertur}), i.e. for the outgoing mode of the
travelling wave solution, it can be seen that the presence of viscosity 
causes the amplitude of the perturbation to diverge exponentially on 
large length scales, where $c_{\mrm{s0}} \simeq c_{\mrm{s}}(\infty)$ 
and $v_0 \sim r^{-5/2}$, with $-v_0$ being positive for inflows.
The inwardly travelling mode also displays similar behaviour, albeit
to a quantitatively lesser degree.
It is an easy exercise to see that stability in the system would be
restored for the limit of $\alpha = 0$, and this particular issue has
been discussed by~\citet{ray03a} and~\citet{crd06}. The exponential growth 
behaviour of the amplitude of the perturbation, therefore, is exclusively 
linked to the presence of viscosity. Going back to a work 
of~\citet{ct93}, it can be seen that the inertial-acoustic modes of short 
wavelength radial perturbations are locally unstable throughout
the disc, with the outward travelling modes growing faster than the
inward travelling modes in most regions of the disc, all of which
is very much in keeping with what equation~(\ref{fpertur}) indicates here.

With the help of equation~(\ref{flucden}) it should be easy to express 
the density fluctuations in terms of $f^{\prime}$ as
\begin{equation}
\label{effden}
\frac{\rho^{\prime}}{\rho_0} = \beta^2 \left(\frac{v_0}{{\mrm i}\omega}
\frac{{\mrm d} s}{{\mrm d} r} \right)\frac{f^{\prime}}{f_0} , 
\end{equation}
and likewise, the velocity fluctuations may be set down from 
equation~(\ref{effprime}) as 
\begin{equation}
\label{effvel}
\frac{v^{\prime}}{v_0} = \left(1 - \frac{v_0}{{\mrm i}\omega}
\frac{{\mrm d} s}{{\mrm d} r} \right)\frac{f^{\prime}}{f_0} . 
\end{equation}

In a unit volume of the fluid, the kinetic energy content is
\begin{equation}
\label{ekin}
{\mathcal E}_{\mrm{kin}} = \frac{1}{2} \left(\rho_0
+ \rho^{\prime}\right)\left(v_0 + v^{\prime}\right)^2 , 
\end{equation}
while the potential energy per unit volume of the fluid is the sum of
the gravitational energy, the rotational energy and the internal energy.
For a quasi-viscous disc, to a first order in $\alpha$, this sum is 
given by
\begin{equation}
\label{epot}
{\mathcal E}_{\mrm{pot}} = \left(\rho_0 + \rho^{\prime}\right) 
\left[ V(r) + \frac{\lambda_{\mrm{eff}}^2}{2r^2} \right]
+ \rho_0 \epsilon + \rho^{\prime} \left[\frac{\prt}{\prt \rho_0}
\left(\rho_0 \epsilon \right)\right] + \frac{1}{2}{\rho^{\prime}}^2
\left[ \frac{{\prt}^2}{\prt \rho_0^2}\left(\rho_0 \epsilon \right) \right] , 
\end{equation}
where $\epsilon$ is the internal energy per unit mass~\citep{ll87}.
In equation~(\ref{epot}) the effective angular momentum for the 
quasi-viscous disc will have to be set up as a first-order correction
about the inviscid conditions. Following this, a time-dependent 
perturbation has to be imposed about the stationary values of $v$ 
and $\rho$. All first-order terms involving time-dependence in  
equations~(\ref{ekin}) and~(\ref{epot}) will 
vanish on time-averaging. In this situation the leading contribution 
to the total energy in the perturbation comes from the second-order 
terms, which are all summed as 
\begin{displaymath}
\label{order2}
{\mathcal E}_{\mrm{pert}} = \frac{1}{2} \rho_0 {v^{\prime}}^2
+ v_0 \rho^{\prime} v^{\prime} +  \frac{1}{2} {\rho^{\prime}}^2
\left[\frac{{\prt}^2}{\prt \rho_0^2} \left(\rho_0 \epsilon \right) \right]
- 2 \alpha \lambda^2 \frac{\rho_0 \sigma}{r^2} \bigg[ \left(
\frac{\rho^{\prime}}{\rho_0}\right)^2 + \left(\frac{f^{\prime}}
{f_0}\right)^2 + 6 \left(\frac{v^{\prime}}{v_0}\right)^2
- 2 \frac{\rho^{\prime}f^{\prime}}{\rho_0 f_0}
+ 3 \frac{v^{\prime} \rho^{\prime}}{v_0 \rho_0} 
- 6 \frac{f^{\prime} v^{\prime}}{f_0 v_0} 
\end{displaymath}
\begin{equation}
\label{contdorder2}
\qquad \qquad \qquad \qquad \qquad \qquad
+ \frac{1}{\sigma} \left(\frac{\rho^{\prime}}{\rho_0} \right) \int 
\sigma \frac{\mrm d}{{\mrm d} r} \left(\frac{f^{\prime}}{f_0} \right)\, 
{\mrm d}r + \frac{1}{\sigma} \int \sigma \left(\frac{f^{\prime}}{f_0}
- 2 \frac{\rho^{\prime}}{\rho_0} - 3 \frac{v^{\prime}}{v_0} \right)
\frac{\mrm d}{{\mrm d} r} \left(\frac{f^{\prime}}{f_0} \right)\,
{\mrm d}r \bigg] . 
\end{equation}
In the preceding expression all terms involving $\rho^{\prime}$ and 
$v^{\prime}$ can be written in terms of $f^{\prime}$ with the help of
equations~(\ref{effden}) and~(\ref{effvel}), in both of which, to a 
leading order, $s \simeq \omega k_{-1}$. This is to be followed 
by a time-averaging over ${f^{\prime}}^2$, which will  
contribute a factor of $1/2$. The total energy flux in the perturbation 
is obtained by multiplying ${\mathcal E}_{\mrm{pert}}$
by the propagation velocity $(v_0 \pm \beta c_{\mrm{s0}})$ and then
by integrating over the area of the cylindrical face of the accretion
disc, which is $2 \pi rH$. Under the thin-disc approximation, $H \ll r$,
this will make it possible to derive an estimate for the energy flux as 
\begin{displaymath}
\label{flux1}
{\mathcal F} (r) \simeq \frac{\pi \beta^2 A_{\pm}^2}{f_0}
\sqrt{\frac{\gamma K}{GM}}
\left[\pm 1 + \frac{1- \beta^2\left(2 - \mu \right)}
{2 \beta \left({\mathcal M} \pm \beta \right)}\right]
\left(1 - \frac{2 \alpha \lambda^2 \psi}
{\beta^2 {\mathcal M}^2 v_0 v_{\mrm K}r^2} \right) 
\end{displaymath} 
\begin{equation}
\label{flux}
\qquad \qquad \qquad \qquad \qquad  \qquad \qquad \qquad 
\qquad \qquad \times
\exp \left[ \pm 2 \alpha \lambda^2 \left(3 \gamma -1 \right) \int
\frac{\beta c_{\mrm{s0}}\left(v_0 \pm \beta c_{\mrm{s0}}\right)}
{v_0 v_{\mrm{K}} r^3\left(v_0^2 - \beta^2 c_{\mrm{s0}}^2\right)}
\, {\mrm d}r \right] , 
\end{equation}
in which ${\mathcal M}$ is the Mach number, as defined earlier, while 
\begin{displaymath}
\label{mu}
\mu = \frac{\rho_0}{c_{\mrm{s0}}^2} \left[
\frac{\prt^2\left(\rho_0 \epsilon\right)}{\prt \rho_0^2} \right] ,  
\end{displaymath}
and 
\begin{displaymath}
\label{exppsi}
\psi = 2 \beta \left[\left(\beta^2 -1 \right)^2 {\mathcal M}^2
\pm {\mathcal M} \beta \left(\beta^2 - 4 \right) + \beta^2 \right]
\left[1 \pm 2 \beta {\mathcal M} + \beta^2 \mu\right]^{-1} . 
\end{displaymath}
When ${\mathcal M} \longrightarrow 0$ on large length scales, $\psi$ 
converges to a finite value. However, on these same length 
scales, what will {\em not} converge are the two terms involving $\alpha$
in equation~(\ref{flux}). Under the asymptotic conditions on $v_0$ and 
$c_{\mrm{s0}}$, discussed earlier, one term will diverge exponentially 
as $r$, while another will have a power-law growth behaviour of $r^6$. 
The quasi-viscous disc will, therefore, be most pronouncedly unstable 
on large scales under the passage of a linearised radially propagating 
high-frequency perturbation. 
It is easy to check that under inviscid conditions, with $\alpha =0$,  
and for an adiabatic perturbation with $\mu =1$, the disc will 
immediately revert to stable behaviour~\citep{ray03a}. 

\section{Concluding remarks}
\label{sec5}

The instability of the quasi-viscous disc raises some doubts, 
the primary one of these being on the possibility of a long-time 
evolution of the disc towards a stationary end. The quasi-viscous 
disc being a dissipative system, i.e. its energy being allowed to 
be drained away from this system, there cannot be any occasion to  
look for the selection of a particular solution, and a selection
criterion thereof, on the basis of energy minimisation, as it can
be done for an idealised inviscid flow, such as the Bondi solution
in spherical symmetry~\citep{bon52,gar79,rb02}. 
This, of course, shall also affect the flow 
rate, which, in this treatment, has been perturbed and has been 
found unstable. As a result of all this the whole system has been 
left without any well-defined criterion by which it could guide 
itself towards a steady state, transonic or otherwise. 

One very important physical role of viscosity in an accretion disc
is that it determines the distribution of matter in the disc. 
The manner in which viscosity redistributes an annulus of matter 
in a Keplerian flow around an accretor is very well known, with 
the inner region of this disc system drifting in because of dissipation, 
and consequently, through the conservation of angular momentum and 
its outward transport, making it necessary for the outer regions of 
the matter distribution to spread even further 
outwards~\citep{pri81,fkr02}. This state of affairs is qualitatively 
not altered in anyway for the quasi-viscous flow, except for the fact 
that with viscosity being very weak in this case, the outward transport 
of angular momentum can be conspicuous only on very large scales. 
It may rightly be conjectured that the instability that develops on 
the large subsonic scales of a quasi-viscous disc is intimately
connected with the cumulative transfer of angular momentum on 
these very length scales. The accumulation of angular momentum in 
this region may create an abrupt centrifugal barrier against any 
further smooth inflow of matter. This adverse effect, on the other 
hand, could disappear if there
should be some other means of transporting angular momentum from 
the inner regions of the disc. Astrophysical jets could readily
afford such a means, insofar as jets actually cause a physical 
drift of angular momentum vertically away from the plane of the 
disc, instead of along it. 

\section*{Acknowledgements}

This research has made use of NASA's Astrophysics Data System. The 
authors express their indebtedness to Rajaram Nityananda and 
Paul J. Wiita for some helpful comments.


\begin{thebibliography}{99}

\bibitem[Abraham et al.(2006)]{abd06}
Abraham, H., Bili\'c, N., Das, T. K.  2006, Classical and Quantum
Gravity, 23, 2371

\bibitem[Abramowicz et al.(1988)]{acls88}
Abramowicz, M. A., Czerny, B., Lasota, J. P., Szuszkiewicz, E.  1988, 
\apj, 332, 646

\bibitem[Abramowicz \& Kato(1989)]{ak89}
Abramowicz, M. A., Kato, S.  1989, \apj, 336, 304

\bibitem[Abramowicz \& Zurek(1981)]{az81}
Abramowicz, M. A., Zurek, W. H.  1981, \apj, 246, 314

\bibitem[Afshordi \& Paczy\'nski(2003)]{ap03}
Afshordi, N., Paczy\'nski, B.  2003, \apj, 592, 354

\bibitem[Artemova et al.(1996)]{abn96}
Artemova, I. V., Bj\"ornsson, G., Novikov, I. D.  1996, \apj, 461, 565 

\bibitem[Barai et al.(2004)]{bdw04}
Barai, P., Das, T. K., Wiita, P. J.  2004, \apj, 613, L49

\bibitem[Bhattacharya \& Bhattacharjee(2005)]{bb05}
Bhattacharya, S., Bhattacharjee, J. K.  2005, Proceedings of the Indian 
National Science Academy, 71A, 1

\bibitem[Bohr et al.(1993)]{bdp93}
Bohr, T., Dimon, P., Putkaradze, V.  1993, J. Fluid Mech., 254, 635

\bibitem[Bondi(1952)]{bon52}
Bondi, H.  1952, \mnras, 112, 195

\bibitem[Chakrabarti(1989)]{c89}
Chakrabarti, S. K.  1989, \apj, 347, 365

\bibitem[Chakrabarti(1990)]{skc90}
Chakrabarti, S. K.  1990, Theory of Transonic Astrophysical
Flows, World Scientific, Singapore

\bibitem[Chakrabarti(1996a)]{c96a}
Chakrabarti, S. K.  1996a, \apj, 464, 664

\bibitem[Chakrabarti(1996b)]{c96b}
Chakrabarti, S. K.  1996b, \apj, 471, 237

\bibitem[Chakrabarti(1996c)]{c96c}
Chakrabarti, S. K.  1996c, \mnras, 283, 325 

\bibitem[Chandrasekhar(1987)]{sc87}
Chandrasekhar, S.  1987, Ellipsoidal Figures of Equilibrium, 
Dover Publications, New York 

\bibitem[Chaudhury et al.(2006)]{crd06}
Chaudhury, S., Ray, A. K., Das, T. K.  2006, \mnras, 373, 146

\bibitem[Chen et al.(1997)]{cal97}
Chen, X., Abramowicz, M. A., Lasota, J. P.  1993, \apj, 476, 61

\bibitem[Chen \& Taam(1993)]{ct93}
Chen, X., Taam, R.  1993, \apj, 412, 254 

\bibitem[Cross \& Hohenberg(1993)]{ch93}
Cross, M. C., Hohenberg, P. C.  1993, Reviews of Modern Physics, 65, 851

\bibitem[Das(2002)]{das02}
Das, T. K.  2002, \apj, 577, 880

\bibitem[Das(2004)]{das04}
Das, T. K.  2004, \mnras, 349, 375

\bibitem[Das et al.(2006)]{dbd06}
Das, T. K., Bili\'c, N., Dasgupta, S.  2006, preprint (astro-ph/0604477)

\bibitem[Das et al.(2003)]{dpm03}
Das, T. K., Pendharkar, J. K., Mitra, S.  2003, \apj, 592, 1078

\bibitem[Frank et al.(2002)]{fkr02}
Frank, J., King, A., Raine, D.  2002, Accretion Power in
Astrophysics, Cambridge University Press, Cambridge

\bibitem[Fukue(1987)]{fuk87}
Fukue, J.  1987, PASJ, 39, 309

\bibitem[Garlick(1979)]{gar79}
Garlick, A. R.  1979, \aap, 73, 171

\bibitem[Goswami et al.(2007)]{gkrd07}
Goswami, S., Khan, S. N., Ray, A. K., Das, T. K.  2007, \mnras 
(To appear), preprint (astro-ph/0703162)

\bibitem[Kafatos \& Yang(1994)]{ky94}
Kafatos, M., Yang, R. X.  1994, \mnras, 268, 925

\bibitem[Kato(1978)]{kato78}
Kato, S.  1978, \mnras, 185, 629

\bibitem[Kato et al.(1988)]{kat88}
Kato, S., Honma, F., Matsumoto, R.  1988, \mnras, 231, 37 

\bibitem[Landau \& Lifshitz(1987)]{ll87}
Landau, L. D., Lifshitz, E. M.  1987, Fluid Mechanics,
Butterworth-Heinemann, Oxford

\bibitem[Liang \& Thomson(1980)]{lt80}
Liang, E. P. T., Thomson, K. A.  1980, \apj, 240, 271

\bibitem[Lightman \& Eardley(1974)]{le74}
Lightman, A. P., Eardley, D. M.  1974, \apj, 187, L1

\bibitem[Livio \& Shaviv(1977)]{ls77}
Livio, M., Shaviv, G.  1977, \aap, 55, 95

\bibitem[Lu et al.(1997)]{lyyy97}
Lu, J. F., Yu, K. N., Yuan, F., Young, E. C. M.  1997, \aap, 321, 665

\bibitem[Matsumoto et al.(1984)]{mkfo84}
Matsumoto, R., Kato, S., Fukue, J., Okazaki, A. T.  1984, PASJ, 36, 71

\bibitem[Molteni et al.(1996)]{msc96}
Molteni, D., Sponholz, H., Chakrabarti, S. K.  1996, \apj, 457, 805

\bibitem[Muchotrzeb-Czerny(1986)]{bmc86}
Muchotrzeb-Czerny, B.,  1986, Acta Astronomica, 36, 1

\bibitem[Nakayama \& Fukue(1989)]{nf89}
Nakayama, K., Fukue, J.  1989, PASJ, 41, 271

\bibitem[Narayan \& Yi(1994)]{ny94}
Narayan, R., Yi, I.  1994, \apj, 428, L13

\bibitem[Nowak \& Wagoner(1991)]{nw91}
Nowak, A. M., Wagoner, R. V.  1991, \apj, 378, 656 

\bibitem[Paczy\'nski \& Wiita(1980)]{pw80}
Paczy\'nski, B., Wiita P. J.  1980, \aap, 88, 23

\bibitem[Pariev(1996)]{par96}
Pariev, V. I.  1996, \mnras, 283, 1264 

\bibitem[Peitz \& Appl(1997)]{pa97}
Peitz, J., Appl, S.  1997, \mnras, 286, 681 

\bibitem[Petterson et al.(1980)]{pso80}
Petterson, J. A., Silk, J., Ostriker, J. P.  1980, \mnras, 191, 571

\bibitem[Pringle(1981)]{pri81}
Pringle, J. E.  1981, \araa, 19, 137

\bibitem[Ray(2003a)]{ray03a}
Ray, A. K.  2003a, \mnras, 344, 83

\bibitem[Ray(2003b)]{ray03b}
Ray, A. K.  2003b, \mnras, 344, 1085

\bibitem[Ray \& Bhattacharjee(2002)]{rb02}
Ray, A. K., Bhattacharjee, J. K.  2002, \pre, 66, 066303

\bibitem[Ray \& Bhattacharjee(2004)]{rb04}
Ray, A. K., Bhattacharjee, J. K.  2004, preprint (cond-mat/0409315)

\bibitem[Shakura \& Sunyaev(1973)]{ss73}
Shakura, N. I., Sunyaev, R. A.  1973, \aap, 24, 337 

\bibitem[Shakura \& Sunyaev(1976)]{ss76}
Shakura, N. I., Sunyaev, R. A.  1976, \mnras, 175, 613

\bibitem[Singha et al.(2005)]{sbr05}
Singha, S. B., Bhattacharjee, J. K., Ray, A. K.  2005, 
Eur. Phys. J. B, 48, 417 

\bibitem[Theuns \& David(1992)]{td92}
Theuns, T., David, M.  1992, \apj, 384, 587

\bibitem[Umurhan et al.(2006)]{unrs06}
Umurhan, O. M., Nemirovsky, A., Regev, O., Shaviv, G.  2006, \aap, 446, 1

\bibitem[Umurhan \& Shaviv(2005)]{us05}
Umurhan, O. M., Shaviv, G.  2005, \aap, 432, L31

\bibitem[Yang \& Kafatos(1995)]{yk95}
Yang, R. X., Kafatos, M.  1995, \aap, 295, 238

\end{thebibliography}
\end{document}